\documentclass[usenatbib,psfig]{mn2e}

\usepackage{epsfig}

\begin{document}


\title[Magellanic satellite galaxies]{On the scarcity of Magellanic Cloud-like satellites}
\author[Phil A. James and Clare F. Ivory]{Phil A. James\thanks{E-mail:
paj@astro.livjm.ac.uk (PAJ); cfi@astro.livjm.ac.uk (CFI)} and Clare F.
Ivory\footnotemark[1]\\
Liverpool John Moores University, Birkenhead, CH41 1LD, UK}

\date{Accepted . Received ; in original form }

\pagerange{\pageref{firstpage}--\pageref{lastpage}} \pubyear{2009}

\maketitle

\label{firstpage}

\begin{abstract}

We have used H$\alpha$ narrow-band imaging to search for star-forming
satellite galaxies around 143 luminous spiral galaxies, with the goal
of quantifying the frequency of occurrence of satellites resembling
the Magellanic Clouds, around galaxies comparable to the Milky Way.
For two-thirds of the central galaxies, no star-forming satellites are
found, down to luminosities and star-formation rates well below those
of the Magellanic Clouds.  A total of 62 satellites is found,
associated with 47 of the central galaxies searched.  The $R$-band
magnitude difference between central galaxies and their satellites has
a median value of 4.6~mag, and a maximum of 10.2~mag.  The mean
projected separation of the satellites from their central galaxies is
81~kpc, or 98~kpc for systems beyond 30~Mpc.  Thus star-forming
satellites are quite rare, and the Milky Way is unusual both for the
luminosity and the proximity of its two brightest satellites.
We also find that the Clouds themselves are unusual in that they appear
to form a bound binary pair; such close satellite pairs, of any luminosity,
are also extremely rare in our survey.

\end{abstract}


\section{Introduction}

Satellite galaxies have attracted an enormous amount of observational
and theoretical study over the past decade.  This is partly a
consequence of the hierarchical nature of galaxy formation in the
currently-popular $\Lambda$CDM models, within which some or all dwarf
satellites may represent left-over building blocks from an earlier
assembly phase of large galaxies.  These models also suffer from the
so-called `substructure problem' \citep{klyp99,moor99}, in that the
predicted numbers of low-mass dark matter haloes are far larger than
the observed numbers of low-luminosity galaxies that might naturally
be expected to occupy them.  The problem has been addressed in the
latest models \citep{bens02,simo07,kopo09} by decreasing or completely
suppressing the star-formation (SF) efficiency in low mass haloes.  One
great success of these models was that they predicted the existence of
ultra-low mass dwarf galaxies substantially fainter than the canonical
dwarf spheroidals.  These ultra-faint dwarfs were subsequently
discovered, primarily through searches of data from the Sloan Digital
Sky Survey (SDSS) \citep{will05,kopo07,wals07}.

Satellites are also potentially of importance in other areas of galaxy
physics.  Merging of low-mass satellites with their central hosts
(minor mergers) is one route for the formation of thick disk
components, and for building bulges.  For example, \citet{domi08} cite
repeated minor merger episodes as their preferred mechanism through
which bulges can grow without destroying the surrounding disk, thus
preserving the strong correlation between disk and bulge colours
within individual galaxies found by these authors.  Another important
problem concerns the continued gas supply to large disk galaxies,
first noted by \citet{lars80}.  These authors estimate that the
Milky Way (MW) will consume the current disk gas reservoir in
$\sim$2~Gyr, and that the equivalent timescale for 36 external
galaxies is little longer, with a mean of 3.9~Gyr, and certainly less than a
Hubble time.  This points clearly to the need for continued gas supply
to support ongoing SF in most or all spiral galaxies.  Such continued
or even increasing supply is also indicated by the K-dwarf metallicity
distribution in the MW, investigated by \citet{casu04}, who also
conclude that the `burstiness' apparent in the SF history of the MW
may indicate significant gas accretion events.  However,
\citet{grce09} estimate the total gas mass in the current populations
of satellites around the MW and M31 (not including the Magellanic
Clouds), finding a total of only 1 - 2$\times10^8$ M$_{\odot}$ for the
MW.  They conclude that this is too little to provide a satisfactory
explanation of the deficiency of low-metallicity dwarf stars in the
Galactic disk (the `G-dwarf problem'), which requires a continued
supply at an average rate of $\sim$1~M$_{\odot}$~yr$^{-1}$ over a
period of 5-7~Gyr.

Much of the recent interest in satellites has concentrated on
extremely low-luminosity dwarf galaxies, which in general can only be
studied within the Local Group.  However, the more massive satellites
are also of substantial importance.  In this paper we focus on
satellites that are near analogues of the two most luminous objects
around the MW, the Large and Small Magellanic Clouds (LMC and SMC,
collectively MC, henceforth) which are characterised by quite high
masses and luminosities (much higher at least than any other galaxies
within several hundred kpc of the Milky Way), ongoing SF indicating
substantial gas reservoirs, and their nearness to us (50 and 60~kpc
for the LMC and SMC respectively).  They also appear to be close to
one another, probably forming a bound pair, with an interaction
history that may be of great importance, both for their SF histories and
for the origin of the Magellanic Stream of H{\sc i} gas.

For comparison with the
satellite galaxies found in the present study, we assume $R$-band
luminosities of 1.6$\times$10$^9$ and 3.7$\times$10$^8$~L$_{\odot}$,
and SF rates of 0.17 and 0.027~M$_{\odot}$~yr$^{-1}$ for
the LMC and SMC respectively.  The latter are derived from the
H$\alpha$ photometry of \citet{kenn86} with magnitude-dependent
extinction corrections derived following \citet{helm04}.  Again for
comparison purposes, we adopt an $R$-band luminosity for the Milky Way
of 1.5$\times$10$^{10}$~L$_{\odot}$; thus to a distant observer, the
Milky Way would appear 9.4 times more luminous than the LMC, a
difference of 2.43~mag.; and 40.5 times more luminous than the SMC, a
difference of 4.02~mag.

One reason for looking for satellites like the LMC and SMC lies in the
recent simulations, discussed above in the context of ultra-faint
dwarfs.  While these models have had great success in accounting for
the luminosity function of low-mass satellites ($10^4 - 10^8$
L$_{\odot})$, most fail to predict the existence of satellites
resembling the Magellanic Clouds in any significant numbers.  For
example, \citet{bens02} find satellites as massive as the LMC in fewer
than 5\% of simulated haloes that harbour MW-like central galaxies.
\citet{simo07} also find that their models, which rely on reionization
to truncate SF in satellite haloes, accurately explain the numbers of
faint satellites but again do not predict MC-like satellites.
\citet{okam10} confirm that the underprediction of the numbers of the
brightest satellites for MW-like systems occurs even in the most
recent $\Lambda$CDM simulations.

$\Lambda$CDM simulations of the satellite populations of MW-like
galaxies by \citet{kopo09} also fail to match the observed luminosity
function of Local Group satellites in their brightest bin, which
comprises the LMC and SMC.  As a result these authors introduce
additional models with parameters tuned to allow the production of
MC-like satellites. Similarly, \citet{krav10} recently produced models
of MW-like haloes where the whole SF efficiency model is tuned to
produce SMC- and LMC-like satellites.

\citet{boyl09} analyse the statistical properties of haloes likely to
host MW-like galaxies in the Millennium-II $\Lambda$CDM simulation,
and determine the probability of hosting a satellite as massive as the
LMC to be 8-25\%, depending on the mass adopted for the MW halo
(larger masses corresponding to higher probabilities).  Similarly,
they find the probabilities of having a second satellite as massive as
the SMC to be in the range 3.3-20\%.  They argue from this analysis that
the existence of the LMC and SMC favours the higher end of the MW halo
mass range.  However, higher mass favours more recent and substantial
mergers in the past history, making the survival of the MW thin disk
harder to understand. This raises obvious questions about how typical
the Magellanic Clouds are.

As one counter-example to the above studies, \citet{libe07} and
\citet{zava08} presented disk galaxy $N$-body/SPH simulations which {\em
  do} predict the existence of massive satellites resembling the
Magellanic Clouds, which they take as confirmation of the validity of
their models.  Again, the strength of this argument depends on whether
the MW system is a special case in this respect.

Magellanic Cloud-like satellites are also important because of their
potentially significant impact on the growth of stellar mass in disk
galaxies.  \citet{hopk08} present simulations of mergers of disk
galaxies with `live' (dynamically interacting) haloes, which show that
MW-mass disks can survive mergers with satellites one-third of their
mass while remaining recognisable disks, while mergers with satellites
of the mass of the LMC have almost no effect, causing only some disk
thickening.  Even this effect is found to be negligible for satellites
less than one-tenth of the central galaxy mass. Their models also
predict a rapid, radial infall of satellites, almost regardless of
their initial impact trajectory.  However it is not clear whether this
prediction is consistent with the multiply-wound satellite trails
around some nearby galaxies, e.g. those around NGC~5907 found by
\citet{mart08}.  \citet{taff03} present similar modelling, but they
focus on the effect of mass loss from the satellite during the
accretion process, which they find to have a substantial effect on
dynamical friction timescales.  Applying this to the LMC, they predict
it is likely to merge with the MW on a $\sim$4~Gyr timescale.
Overall, these simulations show that mergers with MC-like satellites
provide a plausible mechanism for the fast and efficient growth of
disk galaxies, providing of course such satellites exist in
significant numbers.

However, the assumption that our Galaxy is `typical' in any sense,
whilst frequently made, is very risky; this is a sample of one galaxy
that may be unusual in some respects.  An early test of this
assumption was performed by \citet{holm69} who searched for satellite
galaxies around nearby field galaxies using Palomar Sky Atlas
photographic plates. He found that bright satellite galaxies
comparable to the MC are quite rare, at least within $\sim$57~kpc
projected distance from the central galaxies studied.  This conclusion
was supported by \citet{lorr94}, who used higher-quality photographic
material and concluded that a typical spiral central galaxy only has
one satellite brighter than $M_B=$--16.5 within a projected separation
of 375~kpc. \citet{zari93} present a spectroscopic study of 62 nearby
bright spiral galaxies, finding 69 confirmed satellites around 45 of these
central galaxies.  The 69 satellites have absolute $B$ mags between 
--20 and --13, with the bulk of the distribution between --18 and --14.
Nine of the 69 are as bright as or brighter than the LMC, with the 
SMC being close to the median luminosity.

Here we present a study that aims to build on these
results, using narrow-band imaging to search for satellite galaxies
around disk galaxies in the local Universe. In addition to these statistical
goals, a strong motivation for this work is to identify the best
analogues of the MC system, to provide a comparison sample for further
studies of, for example, the origin of the Magellanic Stream. 

Throughout the paper we
assume a Hubble constant of 70~km~s$^{-1}$~Mpc$^{-1}$, although for
the nearer galaxies we adopt distances corrected for Virgo Cluster
infall only, taken from the NASA/IPAC Extragalactic Database (NED).

\section[]{Methods: satellite detection from H$\alpha$ imaging}

\subsection{Motivation for narrow-band imaging searches}

The major problem in assembling catalogues of satellite galaxies is
background or foreground interloper galaxies, which appear to be
companions through line-of-sight projection.  These can be
statistically accounted for by subtracting counts from comparison
fields, as was done by \citet{holm69}, but this is a highly uncertain
procedure due to clustering, and the small `signal' of true
satellites.  Multi-object spectroscopy is the most secure way to
overcome this problem, but it is time consuming given that the input
catalogues are heavily contaminated by background sources, and that
satellites can be faint and of low surface brightness.

Many of these problems can be overcome by using narrow-band imaging,
through a filter with a passband that includes a bright emission line
(H$\alpha$ in the present paper) at the redshift of the central galaxy
or galaxies targeted. This has the advantages that it can be used over
wide fields given current large-format CCDs, each field requires just
two observations (through the narrow-band H$\alpha$ filter, and a
shorter exposure through a broader filter for continuum subtraction),
and only sources with an emission line within the narrower filter
passband will appear in the continuum-subtracted image, thus excluding
almost all background and foreground galaxies due to the redshifting
of the target emission line out of the passband of the filter.  In
addition, H$\alpha$ narrow-band imaging is highly sensitive to even
low levels of star formation; HII regions tend to be clumpy and of
high surface brightness, even in generally LSB galaxies.

The major limitation of this technique is that it only picks up the
star-forming galaxies within the field, and has no sensitivity for
quiescent galaxies such as most elliptical and dwarf elliptical types.
However, the star-forming satellite galaxies are the most relevant
ones if the goal is to identify potential suorces of gas supply to
central galaxies. It is important here to note the conclusion of
recent papers \citep{meur06,jame08b}, which find that almost all
gas-rich galaxies in the local Universe form stars.  Thus to a good
approximation H$\alpha$ selection can be considered equivalent to
selecting of galaxies with significant cold gas content, although we
note some interesting (but apparently rare) cases of galaxies with
high molecular gas contents which have very low SF rates, e.g. for the
galaxies in Stephan's Quintet \citep{appl06, cluv10}.

\subsection{Previous work}

The methods used in the present paper were first tested using data
from the H$\alpha$GS survey \citep{jame04}, with results being
presented in \citet{jame08c}.  We looked at continuum-subtracted
narrow-band H$\alpha$ imaging for 119 central galaxies, finding a
total of 9 probable star-forming satellites. The typical $R$-band
luminosities and H$\alpha$-derived SF rates of the 9 satellites were
similar to those of the SMC, with the LMC being more luminous than all
9 and having a SF rate larger than 8 of the 9. No central galaxy was
found to have more than one star-forming satellite.  Thus the overall
conclusion was that MC-like bright, star forming satellites seem to be
uncommon.  However, it is important to note some significant caveats
with the results presented in \citet{jame08c}.  The CCD imaging used
for H$\alpha$GS had a relatively small areal coverage ($\sim$
10$^{\prime}$ squared) so there are substantial incompleteness
corrections.  Also, the relatively small volume of the local Universe
used to select the central galaxies, and the aim for complete sampling
of dwarf galaxies, resulted in a restricted number of bright central
galaxies.  Thus, for example, only 31 of the central galaxies studied
in \citet{jame08c} are brighter than 10$^{10}$ L$_{\odot}$ and can thus
be considered truly comparable with the MW.

A potentially important additional consideration with the narrow-band
imaging technique is that uncertainties in the continuum subtraction
process can result in spurious objects which do not in fact have any
line emission. Even for those objects where the line emission is real,
the narrow-band imaging alone cannot rule out the possibility of it
being another line, for example [O{\sc
    iii}]$\lambda$5007~\AA\ redshifted into the narrow-band filter
passband.  Finally, even if it is truly the H$\alpha$ line, the width
of the narrow-band filters used here (80 - 100~\AA) means that
galaxies not truly associated with the central galaxy, separated by
distances as large as 30~Mpc assuming pure Hubble expansion, may be
included as potential satellites.  However, spectroscopy of 13 of our
candidate satellites \citep{ivor10} has confirmed that in every case, 
the emission revealed in
the narrow-band imaging was truly H$\alpha$ emission, very close in
recession velocity to that of the central galaxies (a maximum
difference of 263~km~s$^{-1}$, and a mean difference of only
116~km~s$^{-1}$).  Thus we have good confidence that virtually all of
the sources discussed in the present paper are true satellites,
genuinely associated with the central galaxies, with very small
contamination by interlopers.

\section[]{Sample selection and data}

\subsection{Galaxy sample and observations}

This study utilises the same methods as \citet{jame08c}, but a
completely independent dataset, with very different selection
criteria.  H$\alpha$ imaging was obtained for galaxies hosting recent
core-collapse SNe, for a study of the environments immediately
surrounding the locations of the SN \citep{ande08} giving constraints
on progenitor stars.  Selection by core-collapse SN occurrence
naturally weights the sample towards the brightest and most rapidly
star-forming galaxies, making this an excellent complementary sample
to that used in \citet{jame08c}.

The overall galaxy sample is listed in \citet{ande08}, and so we do
not go into detail here on the sample selection or data reduction.  We
use all data taken with the Wide Field Camera (WFC) on the Isaac
Newton Telescope using either filter number 197 (H$\alpha$, central
wavelength 6568~\AA) or 227 (redshifted H$\alpha$, central wavelength
6657~\AA), where the numbers are from the Isaac Newton Group filter database.  
These filters cover all recession velocities from 0 -
6100~km~s$^{-1}$, with a transition point between the two of
$\sim$2400~km~s$^{-1}$. We do not include galaxies from the sample of
\citet{ande08} that were observed using the [S{\sc ii}] filter centred
on 6725~\AA\ (which are too distant for unambiguous detection of faint
satellites) or those observed with the Liverpool Telescope (where the
field of view of the CCD camera is too small for an efficient satellite
search).

\begin{figure}
\includegraphics[width=67mm,angle=270]{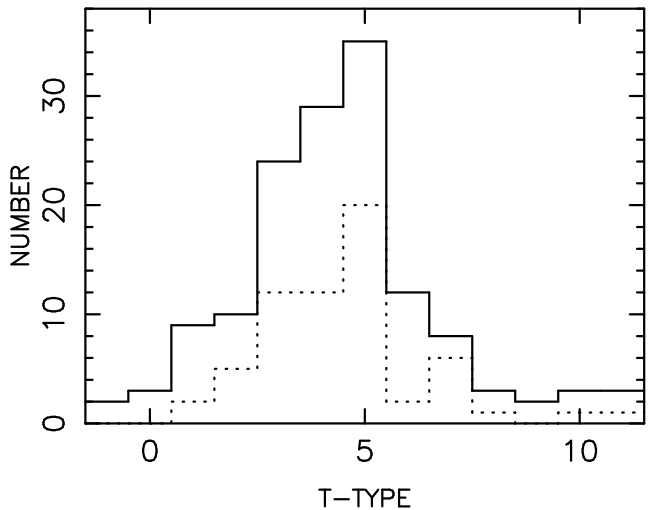}
\caption{Histogram showing the range of Hubble $T$-types of
central galaxies searched for satellites (solid line); and 
the numbers of satellites found, separated by the $T$-types
of their central galaxies.}
\label{fig:thist}
\end{figure}

The remaining dataset of wide-field H$\alpha$ and $R$-band imaging,
that was used to search for star-forming satellites, comprised 143 disk
galaxies, with a distribution of Hubble $T$-types that is
given in Table~\ref{tab_ttype} and shown as the solid line in
Fig.~\ref{fig:thist}. A small number of galaxies with no specific type
classifications in NED were classified by the first author, based on
our own $R$-band images.  Selection by core-collapse supernova
occurrence weights the sample strongly towards bright, star-forming
spirals of types Sb - Scd; these constitute 70\% of the current sample
(100/143).  This closely matches the 66\% contribution of galaxies of
these types to the total star-formation density of the local Universe,
as found by \citet{jame08a}, and thus this appears to be a representative
sample of star-forming galaxies at the present epoch.

\begin{table*}
 \centering
 \begin{minipage}{140mm}
  \caption{The Hubble $T$-type distribution of the parent sample and 
satellite-hosting samples.}
  \begin{tabular}{lrrrrrrrrrrrrrr}
  \hline
   $T$-type~~~ & $<$0 & 0 & 1 & 2 & 3 & 4 & 5 & 6 & 7 & 8 & 9 & 10 & Other\\
   Classn.  & S0   & S0a & Sa & Sab & Sb & Sbc & Sc & Scd & Sd & Sdm & Sm 
            & Im & S, Pec etc & All\\
\hline            
   N$_{CENT}$ tot & 2 & 3 & 9 & 10 & 24 & 29 & 35 & 12 & 8 & 3 & 2 & 3 &  3 & 143\\
   N$_{CENT}$ sat & 0 & 0 & 2 & 4 &  8 & 12 & 12 &  1 & 5 & 1 & 0 & 1 &  1 & 47\\
   N$_{SAT}$      & 0 & 0 & 2 & 5 & 12 & 12 & 20 &  2 & 6 & 1 & 0 & 1 &  1 & 62\\
\hline
\end{tabular}
\label{tab_ttype}
\end{minipage} 
\end{table*}

\subsection{Data reduction}
\label{ssec:datred}

All CCD images used for the satellite detection were bias and dark
subtracted, flat fielded using twilight sky flats, and continuum
subtracted, all using standard procedures as outlined in
\citet{ande08}.  Given the passband of the filters used, it should be
noted that throughout this paper, `H$\alpha$' in fact refers to both
H$\alpha$ and [N{\sc ii}] 6548 \& 6584~\AA\ emission; no attempt is
made to correct for the latter.  The satellite search was done by eye,
scanning the continuum-subtracted H$\alpha$ images for clear
emission-line objects, and `blinking' such objects with the
spatially-registered $R$-band images to confirm that there was a
coincident source with detectable emission in the broader filter
(noting that the latter includes H$\alpha$).  The scanning of images
for satellites was done independently by both authors. Thus for the
present paper, a `satellite' is an object with emission in both
images, that is separated from the main body of the central galaxy.
Given the aims of the present paper, any companions with an $R$-band
luminosity more than one-third of that of the central galaxy were not
included in our analysis or statistics. However, it is important to
bear in mind that there is no universal definition of `satellite' when
comparing different studies, many of which do include such bright
companion galaxies.  Only the central CCD (number 4) of the WFC 4-chip
array was used for the satellite search, to avoid complications caused
by the gaps between the CCDs, and vignetting losses affecting the
outer corners of the array.  The area of the single CCD is $\sim$
12$^{\prime}\times$24$^{\prime}$, corresponding to 111$\times$221~kpc
at the median distance (31.7~Mpc, 2220~km~s$^{-1}$) of the galaxies
searched in this study, or $>$ 190$\times$379~kpc for the galaxies in
the upper quartile of distance.  Thus incompleteness is very small for
satellites as close to their central galaxies as are the Magellanic
Clouds to the Milky Way (50 \& 60 kpc), but a significant fraction of
satellites at the virial radii of the host systems (typically 200 -
300~kpc) will be missed.

This incompleteness can be quantified to some extent using the method
outlined in \citet{jame08c}.  Briefly, this involves calculating the
fraction of a circular volume centred on each of the central galaxies
that would be visible, with the `invisible' regions being both those
lying off the CCD, and those projected onto the disk of the central
galaxy.  These fractions can then be summed for the whole sample and
divided by the number of central galaxies to give an estimated
completeness fraction, under the assumption of uniformly distributed
satellites.  This was first done for the present sample using the
`Magellanic radius' defined in \citet{jame08c}; this is the radius
(76~kpc) corresponding to a sphere with twice the volume necessary to
include both Magellanic Clouds.  Overall, we sample about 75\% of this
volume for the 143 galaxies in the current sample, with the most
substantial loss being for galaxies closer than 20~Mpc.  Beyond 30
Mpc, we see the full volume out to the Magellanic Radius, apart from
the cylinder comprising about 5\% of the total volume that is
projected against the central galaxy disk.  For greater
central-satellite separations, the incompleteness obviously becomes
larger; repeating the above calculation for a radius of 200~kpc, we
find that we can only see 34\% of the corresponding volume.  However,
the $\Lambda$CDM models of \citet{kopo07} predict that only a very
small fraction of satellites lies as far out as 200~kpc, and that
between 70 and 90\% (depending on the model parameters used) lie
within 100~kpc of the simulated central galaxy.  Our survey efficiency
for a 100~kpc radius sphere is 64\%.  Overall, we conclude that we
should pick up about 50\% of all satellites, and 75\% of those within
the `Magellanic radius'.

\section{Results of the satellite search}
\label{sec:res}

The main results of the satellite search are summarised in
Table~\ref{tab_ttype}.  The total number of central galaxies searched was 143,
approximately one-third of which were found to have at least one associated
line-emitting galaxy.  Of these 47 central hosts, 37 have only a single
satellite, 7 have two satellites, 2 have three satellites and one
(NGC~3074) has five satellites; thus the 47 central galaxies host a
total of 62 satellites.  The $T$-types of the central galaxies for
each of the 62 is shown as the dotted line in Fig.~\ref{fig:thist};
this matches the type distribution for all galaxies searched within
statistical uncertainties, i.e. there is no evidence for any galaxy
type having systematically more or fewer satellites than the average.

The mean and median recession velocities of
the satellite-hosting galaxies are 2938 and 2543~km~s$^{-1}$
respectively, somewhat larger than the corresponding values for the
whole sample of 143 galaxies (2539 and 2217~km~s$^{-1}$). This indicates
that the dominant incompleteness is likely to be due to the smaller effective
search area for the nearer systems, rather than missing faint
satellites at larger distances.  In any case, the recession velocity
distributions of the overall and satellite-hosting samples are only
marginally different, with a Kolmogorov-Smirnov test showing a 21\%
chance that they could be drawn from the same parent distribution.

Our $R$-band imaging and the catalogued recession velocities were used
to determine the total luminosities of the central hosts.  The mean
value for the 47 satellite-hosting galaxies was 1.53$\times$10$^{10}$
L$_{\odot}$, almost identical to our adopted luminosity for the MW,
thus confirming that this is a good sample for deriving satellite
properties of MW-like central galaxies.

\begin{figure}
\includegraphics[width=67mm,angle=270]{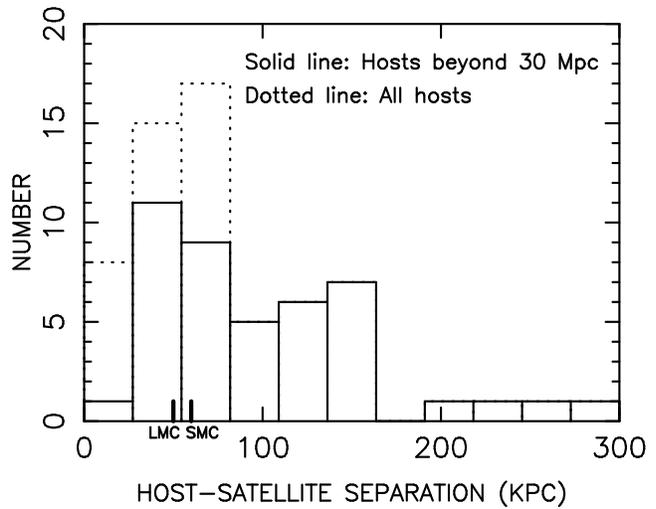}
\caption{Histogram showing the projected separations of the
satellites from the central galaxies.  The dotted line is for the
full sample of 62 satellites, the solid line is for the subset
beyond 30~Mpc.  Thus central galaxies within 30~Mpc only 
contribute close-in satellites, lying less than $\sim$80~kpc in
projected separation from 
their central galaxies. The two thick vertical lines show the actual 
separations of the MCs from the MW.}
\label{fig:satsep}
\end{figure}

\begin{figure}
\includegraphics[width=67mm,angle=270]{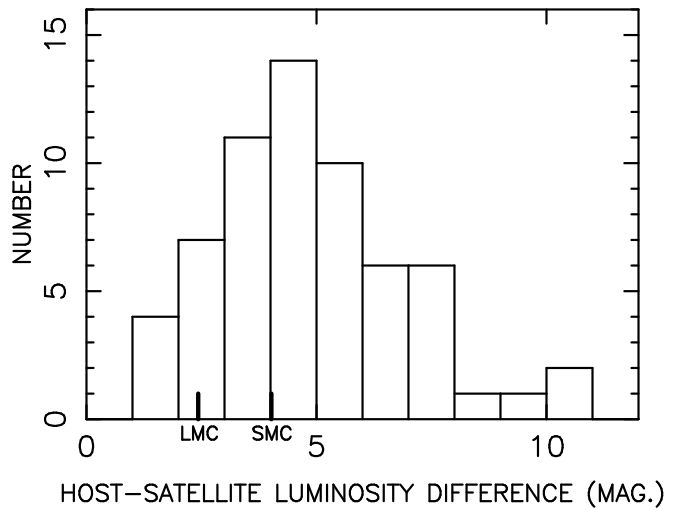}
\caption{Histogram showing distribution of $R$-band magnitude 
differences between satellite galaxies and their central 
host galaxies.  The two thick vertical lines mark the
equivalent differences for the Magellanic Clouds, 
relative to the Milky Way. }
\label{fig:rmagdiff}
\end{figure}

We now move to a consideration of the properties of the satellites that
were detected, with the principal aim in the present paper of identifying those
that can be considered similar to the LMC or SMC.  A more detailed discussion
of, for example, SF rates, evidence for starbursts or suppressed
SF, continuum and emission-line morphologies, and the overall satellite
galaxy luminosity function will be presented
in a future paper (C. F. Ivory et al., in preparation).

The first satellite property to consider is projected separation,
measured from the $R$-band centroid of the host to that of the
satellite.  Figure~\ref{fig:satsep} shows the distribution of these
values for the 62 satellites, with the dotted line corresponding to
all satellites, and the solid line to just those with central galaxies
more distant than 30~Mpc.  The latter thus excludes those systems with
significant incompleteness for outlying satellites due to the smaller
projected area surveyed.  The mean and median projected separations of
the satellites from their central galaxies are 81 and 66~kpc
respectively, rising to 98 and 90~kpc for systems beyond 30~Mpc.
These are of course lower limits to the true separations since the
line-of sight separation is almost completely unconstrained and hence
is not included in these estimates.  The thick vertical lines
correspond to the {\em actual} distances of the LMC and SMC from the
MW; even without correcting for projection effects, it is clear that
the MC should be considered nearby companions, although we do find
other examples with comparable separations.

\begin{figure}
\includegraphics[width=67mm,angle=270]{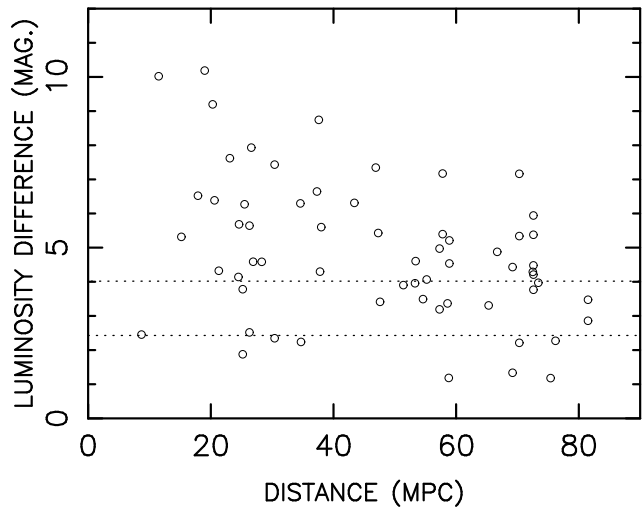}
\caption{$R$-band magnitude
differences between satellite galaxies and their central 
host galaxies plotted against distance in Mpc.  
The horizontal lines show the equivalent magnitude
differences for the LMC (lower) and SMC (upper), illustrating
that we are sensitive to systems with more extreme luminosity 
contrasts than the SMC and Milky Way to the limits of our survey.}
\label{fig:mdiffdist}
\end{figure}

\begin{figure}
\includegraphics[width=67mm,angle=270]{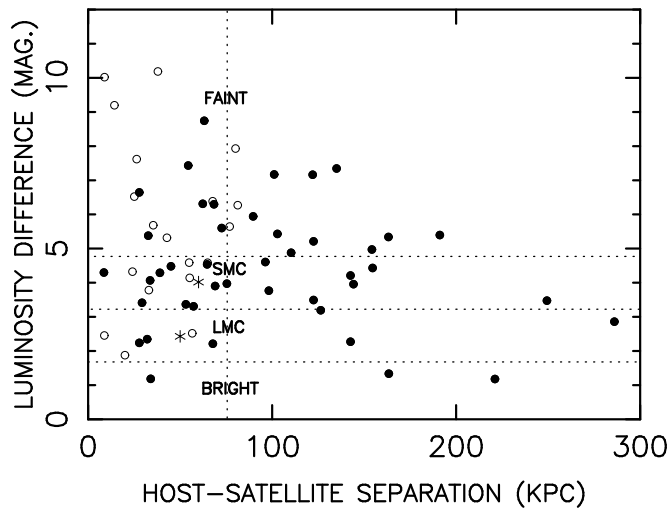}
\caption{$R$-band magnitude
differences between satellite galaxies and their central 
host galaxies plotted against projected separations.  
Open circles are for systems within 30~Mpc, filled
circles lie beyond this distance, and the asterisks show
the estimated values of the Magellanic Clouds (plotted
at their actual distances from the Milky Way). The dashed
lines separate different luminosity and projected separation classes
of satellites, as defined and explained in the text.}
\label{fig:sepmagdiff}
\end{figure}

The next parameter considered is satellite luminosity.  This is
presented in Fig.~\ref{fig:rmagdiff}, where we follow \citet{zari93}
in plotting the difference between the total $R$-band magnitudes of
satellites and their central hosts.  This quantity is chosen as best
representing the evolutionary importance of the satellites in terms of
the dynamical impact of future mergers and their importance as sources
of future gas supply.  It is also a useful quantity for comparison with
cosmological simulations, where luminosity ratios are more easily calculated
than absolute luminosities. Again the equivalent properties are calculated
for the LMC and SMC, using the numbers quoted in the Introduction, and
indicated in Fig.~\ref{fig:rmagdiff} by the thick vertical lines.  In
terms of this relative luminosity, the MC must clearly be considered
very substantial satellites, with the LMC being brighter than 54 of
the 62 satellites found, and the SMC brighter than 40 of the 62. The
distribution shown in Fig.~\ref{fig:rmagdiff} shows a monotonic rise
towards relatively fainter satellites down to those 5~mag fainter than
their central galaxies, followed by a turnover and a tail to the
distribution extending to $\sim$10~mag (showing that we can detect 
satellites with
luminosities as low as 10$^{-4}$ times that of their central galaxy).  To
investigate the effect of incompleteness on this turnover, in
Fig.~\ref{fig:mdiffdist} the same luminosity difference parameter is
plotted as a function of the distance of the galaxy system.  While the
very faintest satellites are only found within $\sim$20~Mpc,
Fig.~\ref{fig:mdiffdist} shows that satellites with a luminosity
difference of $\sim$7~mag are found to virtually the full depth of our
sample.  This is well beyond the turnover at 5~mag, implying that this
turnover is likely to be real, and importantly for the present analysis
shows that we are very unlikely to miss satellites as bright as either 
of the MCs.  The distribution in Fig.~\ref{fig:rmagdiff} is fairly
similar to that found by \citet{zari93} for their sample of 69 
satellites; they find the magnitude difference distribution to rise to
a peak at 3 - 3.5~mag, with a tail out to $\sim$7.5~mag.

Figure~\ref{fig:sepmagdiff} combines the two parameters discussed
above, host-satellite luminosity difference and host-satellite spatial
separation, in one scatter plot that enables an overall comparison of
the 62 satellites with the SMC and LMC, the latter being plotted as
asterisks.  The dotted lines are chosen, somewhat arbitrarily, to
indicate satellites that might be considered most closely analogous to
the SMC and LMC.  The top and bottom horizontal lines correspond to
satellites with one-half of the normalised luminosity of the SMC, and
double the normalised luminosity of the LMC, respectively. The central
line corresponds to both double the SMC normalised luminosity, and
half the LMC luminosity (noting that within errors, the LMC is four
times as bright as the SMC).  Thus the horizontal lines define four
luminosity difference ranges that we term `Bright', `LMC-like',
`SMC-like' and `Faint', as labeled on Fig.~\ref{fig:sepmagdiff}.  The
vertical dashed line corresponds to the `Magellanic radius' defined
above.  Some conclusions are apparent from this figure.  Relative to
these boundaries, three satellites are {\em significantly} brighter
than the LMC; two of these (in the bottom right-hand box) are at
substantial projected distances from their host galaxies and might
better be consider as binary companions or group members. All three of
these bright satellites have luminosities close to one-third that of
their hosts.  There are 9 satellites within a factor 2 of the
normalised LMC luminosity, of which 3 are distant companions but 6 are
sufficiently close in projected separation to be considered `MC-like'.
Similarly, there are 22 satellites within a factor 2 of the normalised
SMC luminosity, with 7 being distant companions and 15
`MC-like'. Finally, 28 of the 62 satellites are more than a factor two
fainter than the SMC in normalised luminosity.  So overall, our search
of 143 bright, star-forming disk galaxies has yielded 21 satellites
which we consider to be similar to the MCs in terms of normalised
luminosity and proximity to their central galaxies; of these, 6 most
closely resemble the LMC, and 15 the SMC.

\section{Multiple satellite systems}

One further distinctive characteristic of the MCs is their binary
nature.  Not only are there two of them, but they are close together
in space, and they appear, from their relative velocities
\citep{krou97}, to be a bound pair. \citet{fuji99} estimate that they
are orbiting with a separation varying between $\sim$ 5 and 30~kpc,
currently $\sim$22~kpc, i.e. substantially less than their separation
from the MW. In this section we look at those systems from the present
sample with more than one star-forming satellite, to see whether there
are any other examples of such binary satellites.

\begin{figure}
\includegraphics[width=85mm,angle=0]{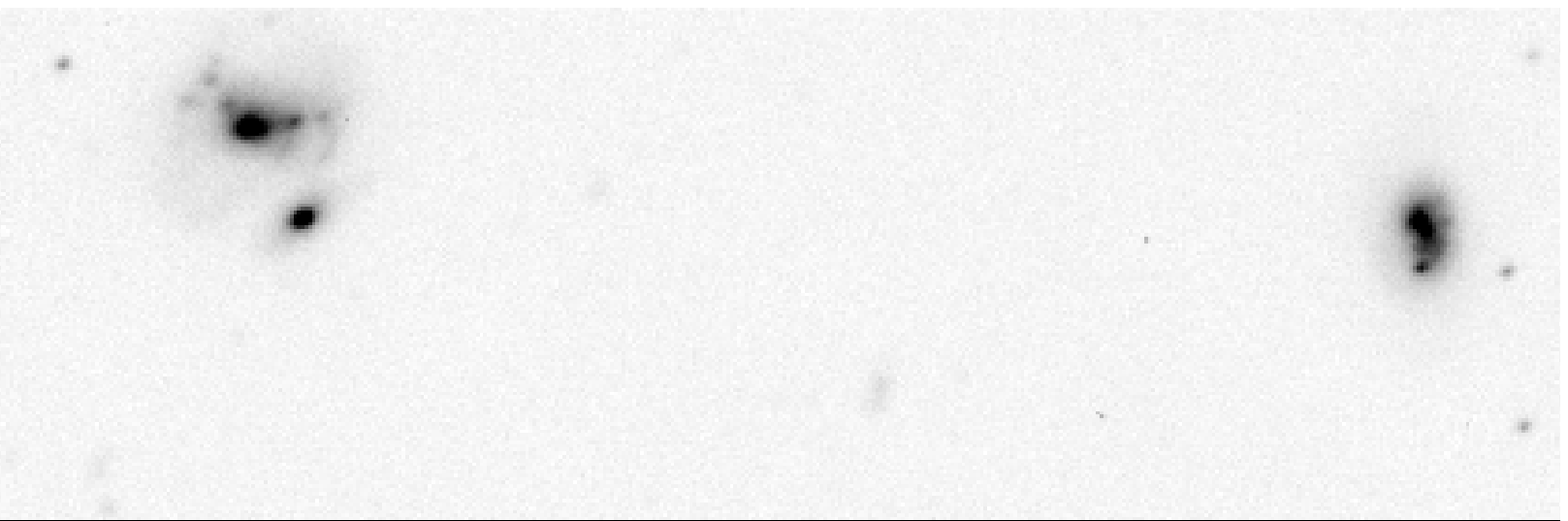}
\includegraphics[width=85mm,angle=0]{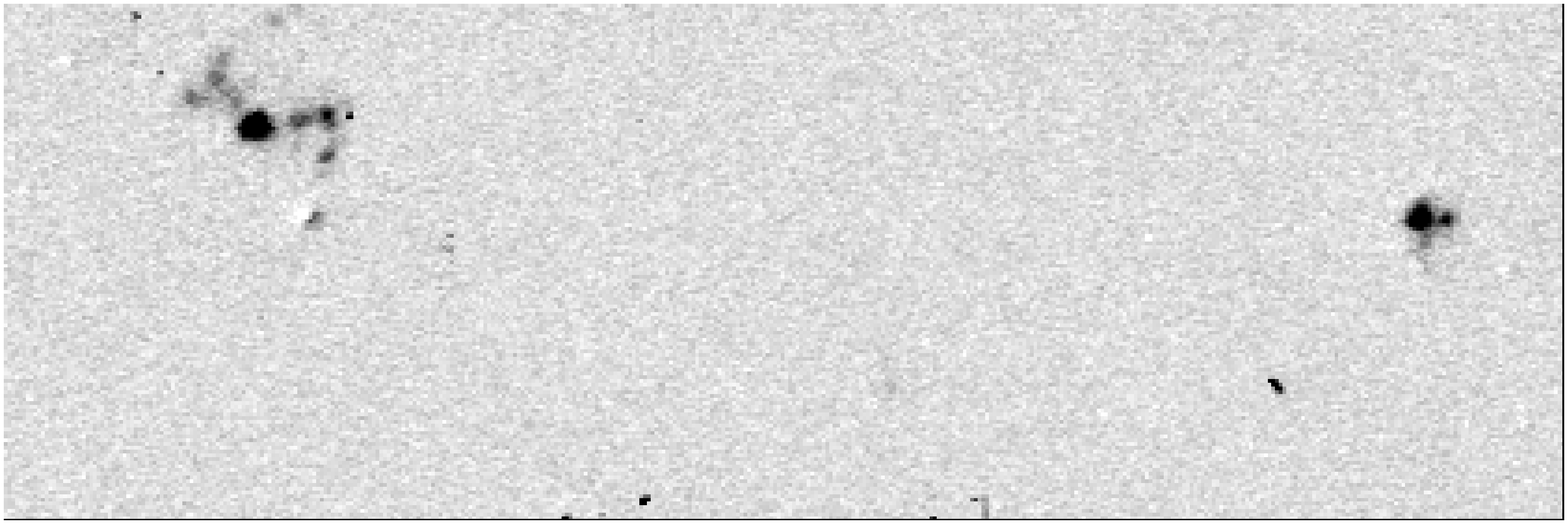}
\caption{$R$-band (top) and H$ \alpha$ (bottom) images of the  
two satellites of NGC~2596.}
\label{fig:n2596}
\end{figure}

\begin{figure}
\includegraphics[width=85mm,angle=0]{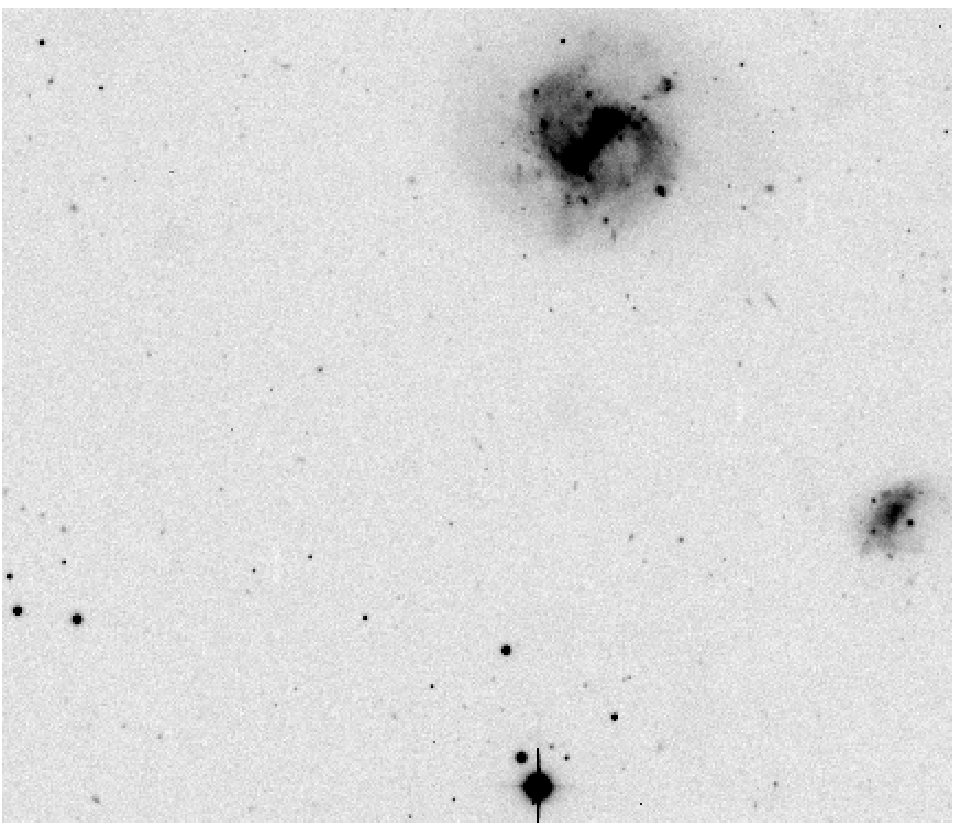}
\includegraphics[width=85mm,angle=0]{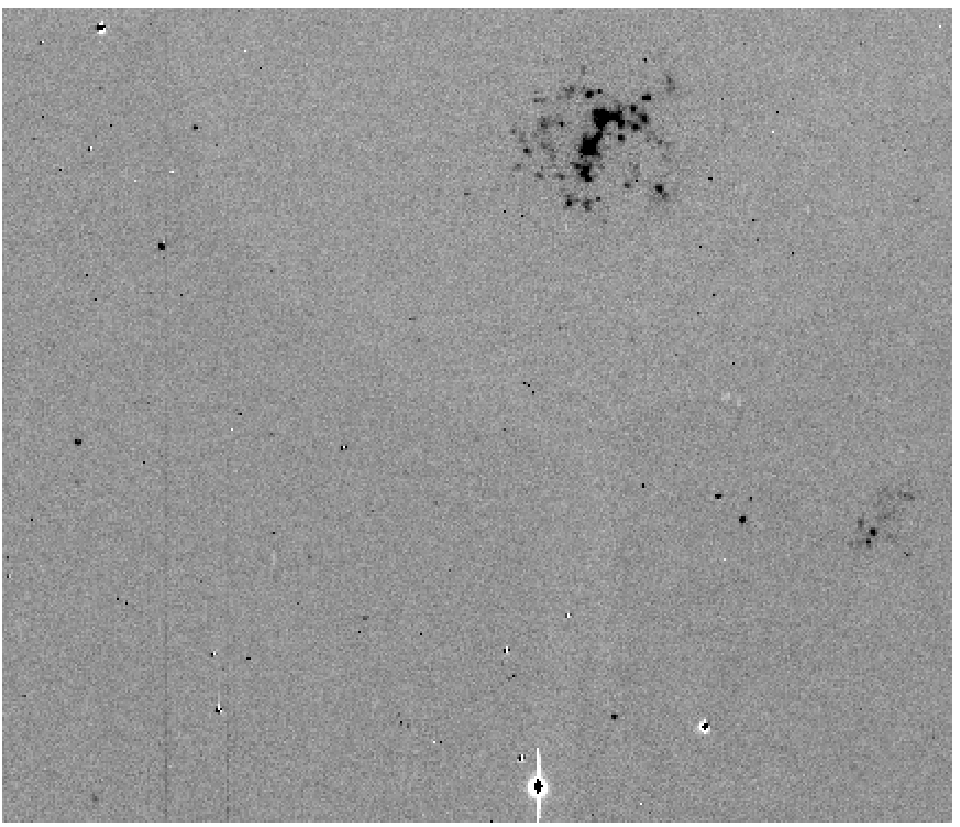}
\caption{$R$-band (top) and H$ \alpha$ (bottom) images of NGC~2604 and 
two satellites. The fainter satellite (barely visible in this reproduction) 
is in the extreme lower-left hand corner of the frames.}
\label{fig:n2604}
\end{figure}

There are 10 host galaxies with 2 or more satellites in the present
sample.  These are listed below, with details of the numbers of
star-forming satellites, the radial projected separation from the
host and the magnitude difference relative to the host for each
satellite, and comments on any pairs of satellites that lie close
to one another in projection.\\

\noindent
{\bf UGC~2627}\\ 
3 satellites found, with the following properties:\\ 
65~kpc separation/4.5~mag difference; 122~kpc separation/5.2~mag difference; 
34~kpc separation/1.2~mag difference.\\

\noindent
This galaxy has one SMC-like companion, one faint and distant
companion, and one close-in and very bright companion that we classify
as too bright to be considered LMC-like.  The bright and SMC-like
companions appear fairly close to one another, with a projected
separation of 40~kpc, but do not appear to be a true pair.\\

\noindent
{\bf NGC~1961}\\
2 satellites found, with the following properties:\\  
126~kpc separation/3.2~mag difference; 154~kpc separation/5.0~mag difference.\\
 
\noindent
The first of these satellites lies on the SMC/LMC mag difference
borderline, the other is faint. Both are distant from the central
galaxy, but closer to one another (42~kpc in projected separation).\\

\noindent
{\bf UGC~4195} \\
3 satellites found, with the following properties:\\ 
68~kpc separation/2.2~mag difference; 122~kpc separation/7.2~mag difference; 
163~kpc separation/5.3~mag difference.\\ 

\noindent
The first of these is a good LMC analogue, the other two are faint and
distant satellites. The two faint satellites are quite close to one another,
41~kpc in projected separation.\\

\noindent
{\bf NGC~2596}\\
2 satellites found, with the following properties:\\ 
286~kpc separation/2.9~mag difference; 249~kpc separation/3.5~mag difference.\\

\noindent
In terms of mag difference, these satellites resemble the LMC and SMC
respectively, but they are the two most distant satellites in the
present sample.  Interestingly, they are quite close to one another,
with a projected separation of 41~kpc. $R$-band and continuum-subtracted 
H$\alpha$ images of the two satellites only (NGC~2596 is not shown)
are presented in Fig.~\ref{fig:n2596}.  Both satellites have clumpy
H$\alpha$ emission with strong central components.\\

\begin{figure}
\includegraphics[width=85mm,angle=0]{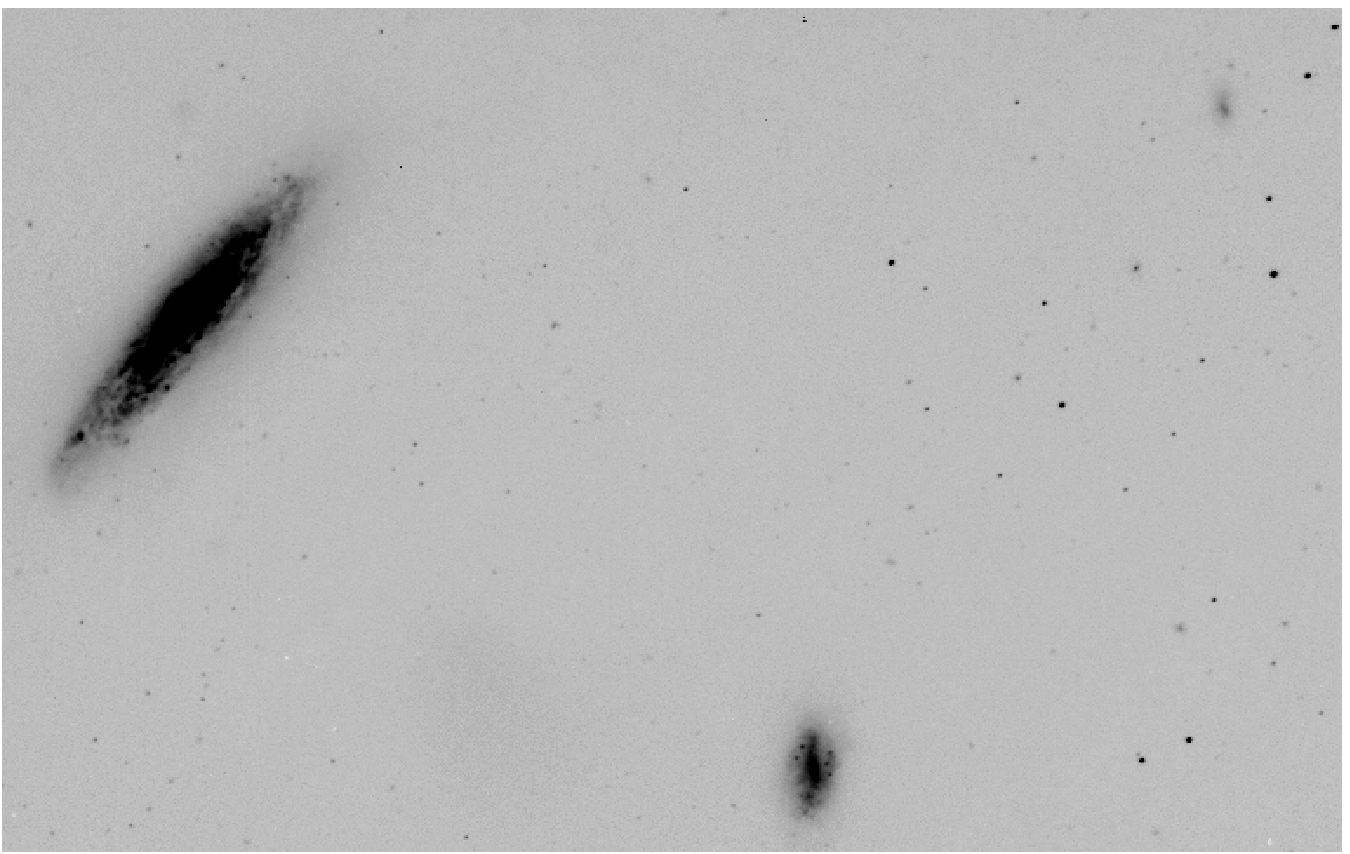}
\includegraphics[width=85mm,angle=0]{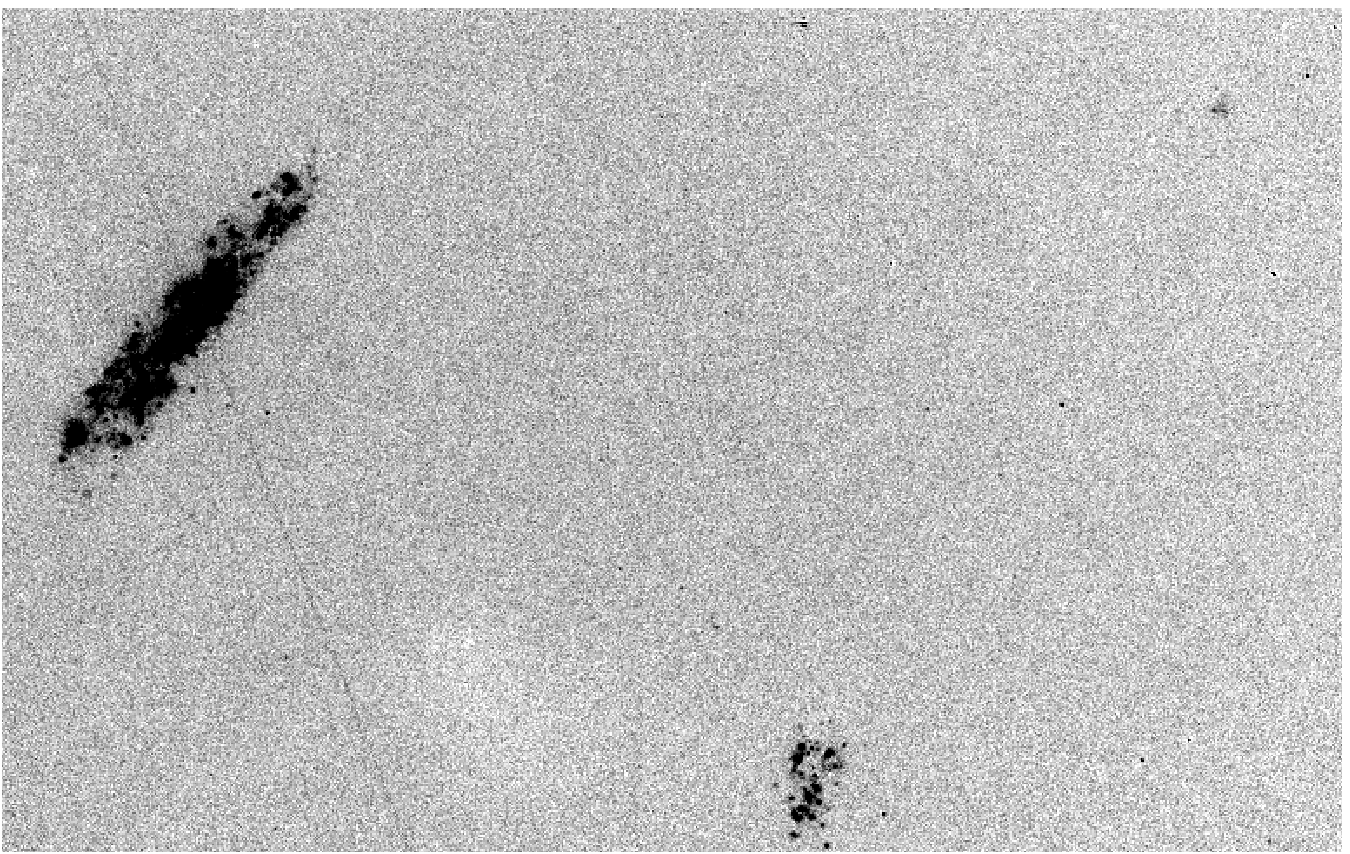}
\caption{$R$-band (top) and H$ \alpha$ (bottom) images of NGC~4666 and 
two satellites. The fainter satellite is in the extreme top-right hand
corner of the frames.}
\label{fig:n4666}
\end{figure}

\noindent
{\bf NGC~2604}\\ 
2 satellites found, with the following properties:\\ 
32~kpc separation/2.3~mag difference; 54~kpc separation/7.4~mag difference.\\ 

\noindent
The first of these is a good LMC analogue (with barred, irregular
structure and multiple H{\sc ii} regions, see Fig.~\ref{fig:n2604}),
the second is very faint but still close to the central galaxy. They
do not appear to be a satellite pair, with a projected separation
between the two greater than the central - satellite separations.\\

\noindent
{\bf NGC~3074}\\ 
5 satellites found, with the following properties:\\ 
33~kpc separation/5.4~mag difference; 45~kpc separation/4.5~mag difference; 
98~kpc separation/3.8~mag difference; 90~kpc separation/5.9~mag difference;  
143~kpc separation/4.2~mag difference.\\ 

\noindent
Three of the five can be considered SMC analogues in terms of mag difference,
and one of these also lies within the `Magellanic radius'. Of the other
two, both are very faint, with one being close in and the other very distant.
In terms of satellite-satellite separations, it should be noted that the two
closest-in satellites (one faint, one SMC-like) are also close to one another,
with a projected separation of 13~kpc. Otherwise, the satellites are widely
spaced.\\

\noindent
{\bf NGC~4666}\\ 
2 satellites found, with the following properties:\\ 
57~kpc separation/2.4~mag difference; 77~kpc separation/5.6~mag difference.\\

\noindent
The first of these is an excellent LMC analogue (see
Fig.~\ref{fig:n4666}); as with the brighter satellite of NGC~2604, it
appears barred in the red continuum image, but irregular with multiple
H{\sc ii} regions in the H$\alpha$ image. The other satellite is very faint, and
quite distant from both the central galaxy and the bright satellite.\\

\begin{figure}
\includegraphics[width=85mm,angle=0]{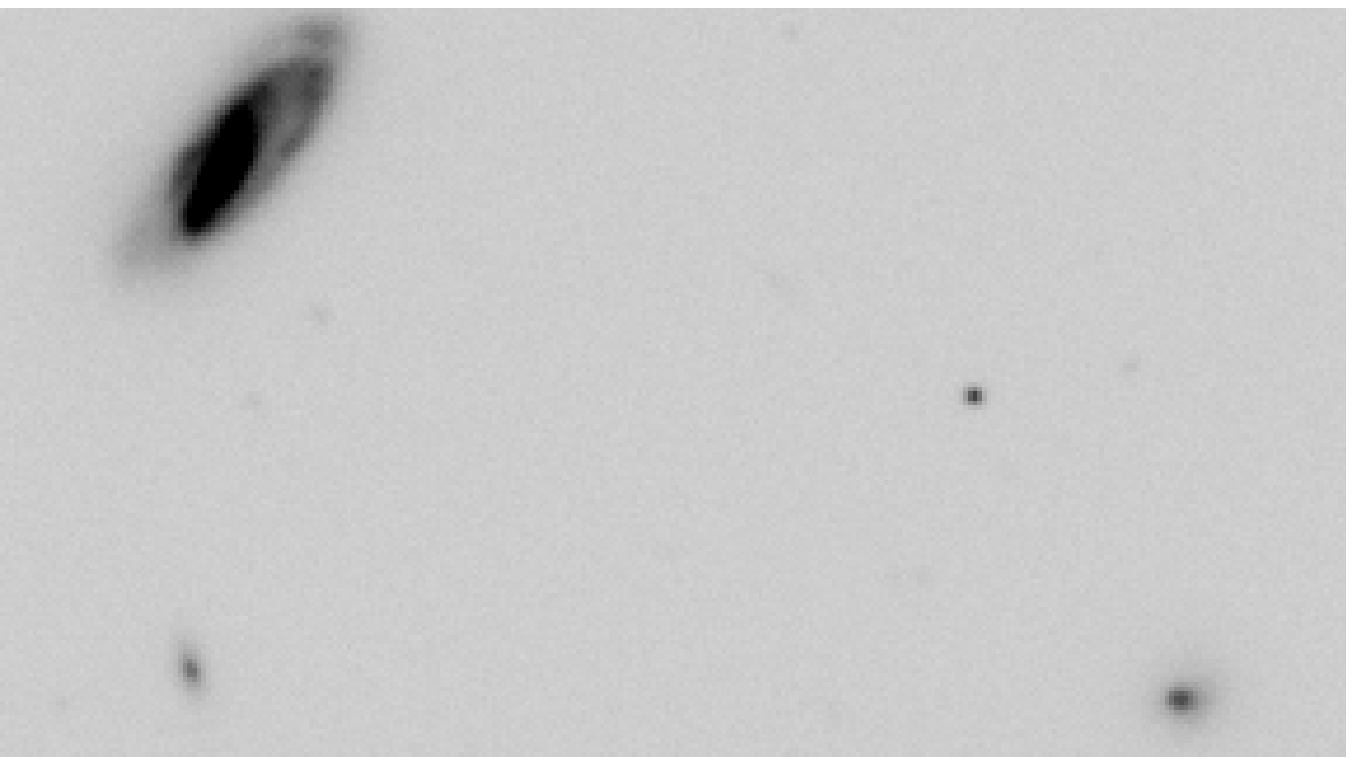}
\includegraphics[width=85mm,angle=0]{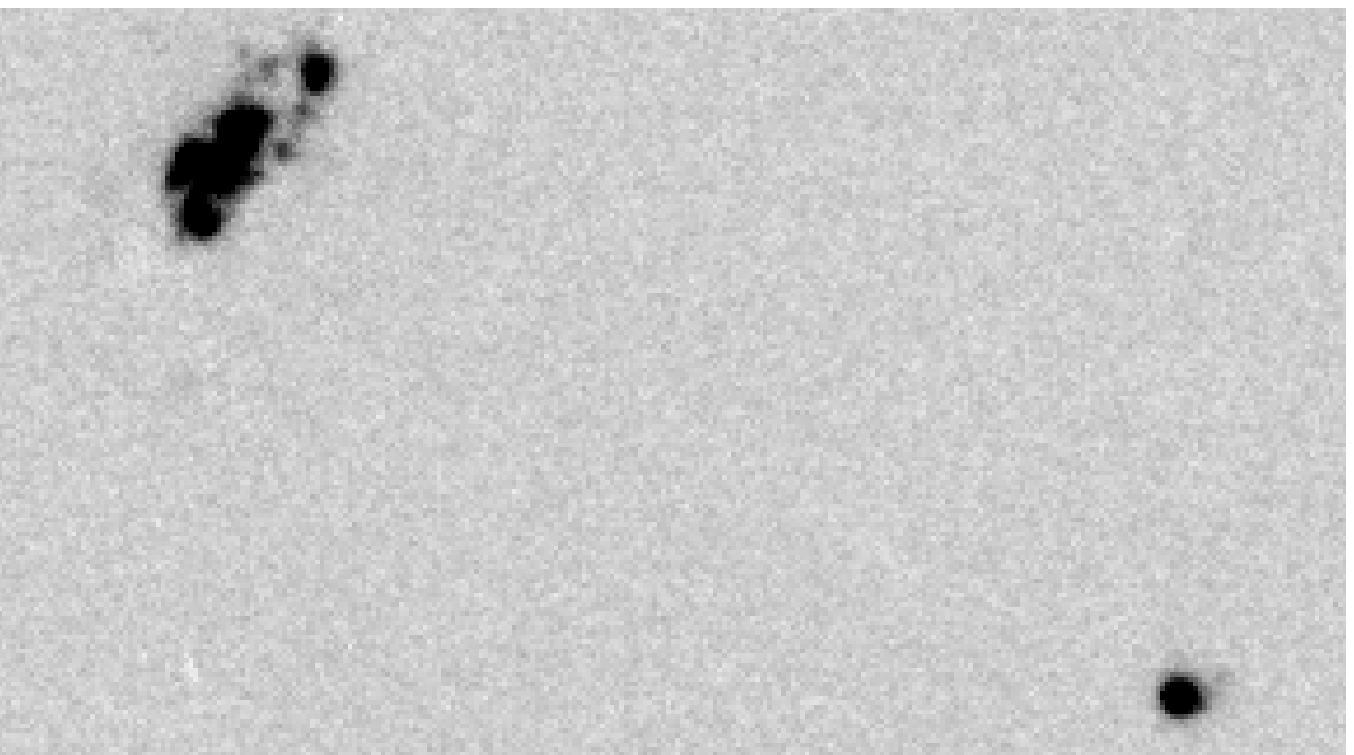}
\caption{$R$-band (top) and H$ \alpha$ (bottom) images of the  
two satellites of NGC~4675.}
\label{fig:n4675}
\end{figure}

\begin{figure}
\includegraphics[width=85mm,angle=0]{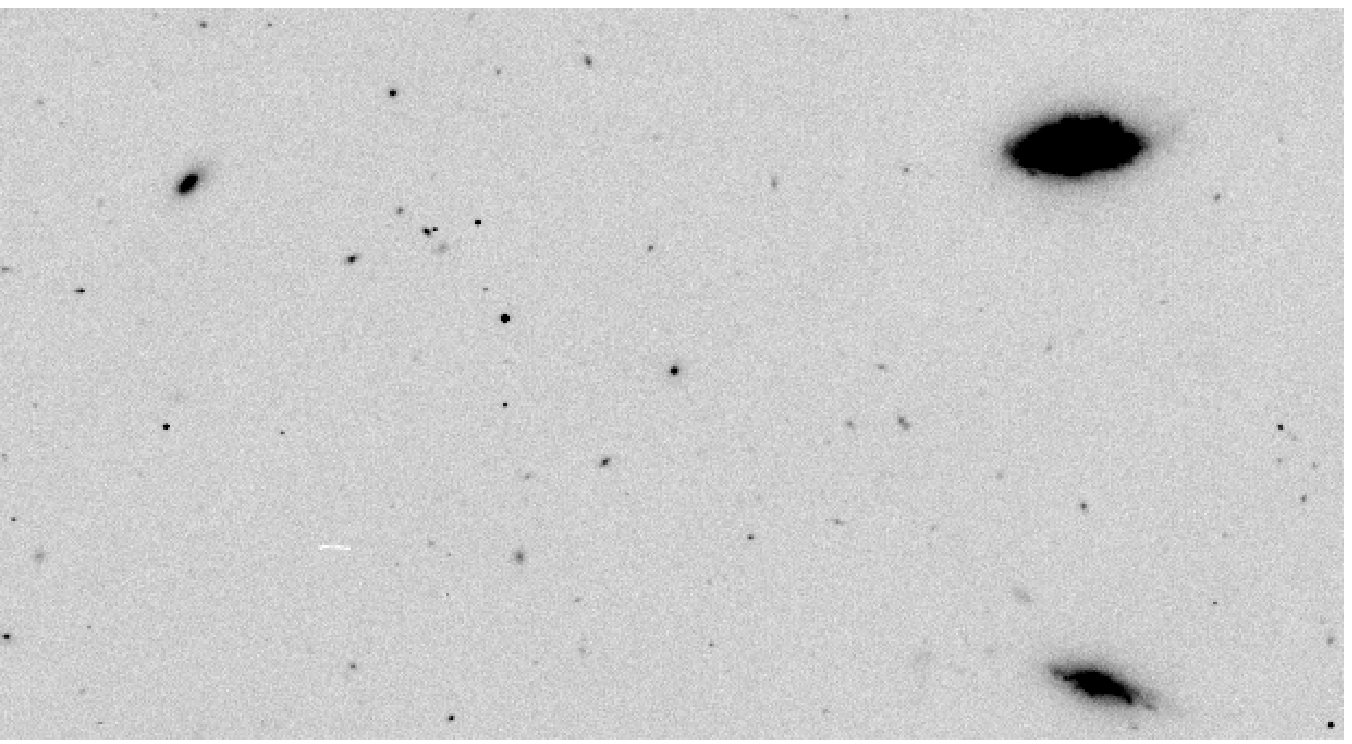}
\includegraphics[width=85mm,angle=0]{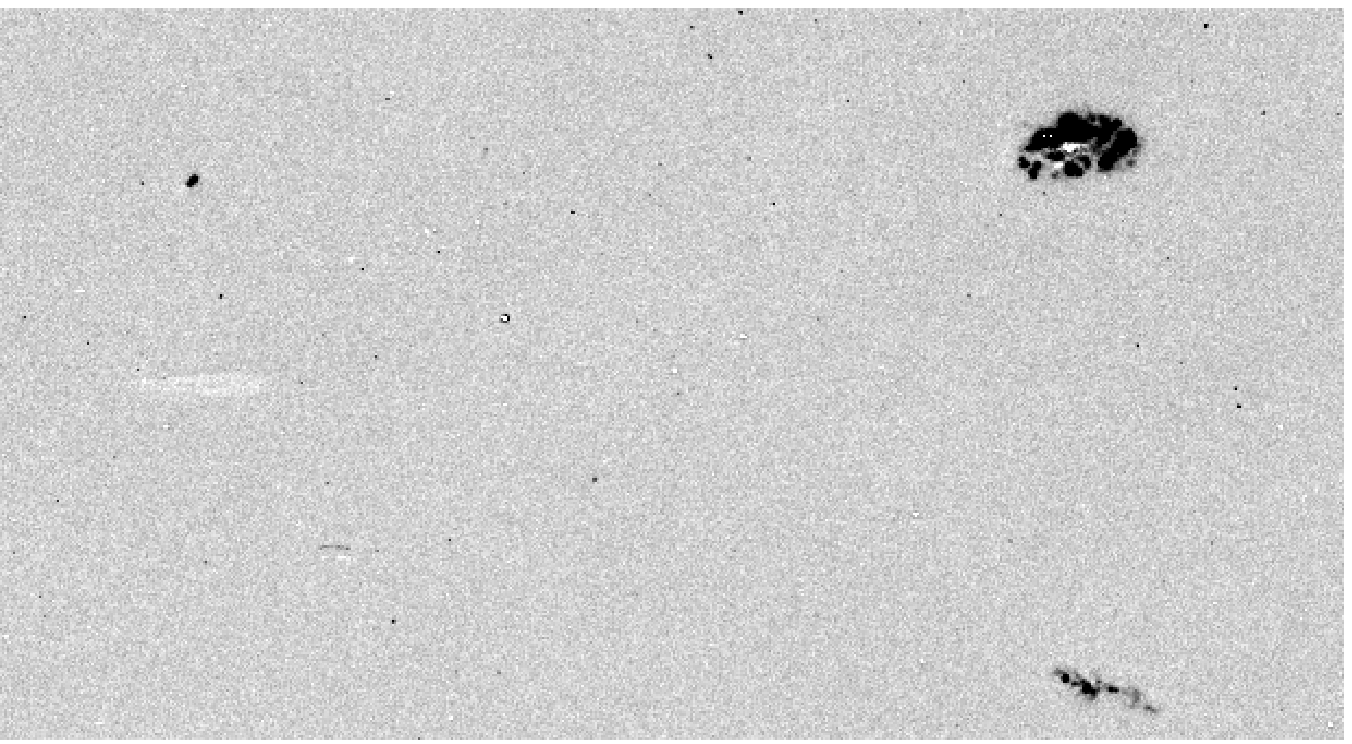}
\caption{$R$-band (top) and H$ \alpha$ (bottom) images of NGC~5303 and
two satellites. The fainter satellite is toward the top-left hand
corner of the frames.}
\label{fig:n5303}
\end{figure}

\noindent
{\bf NGC~4675}\\ 
2 satellites found, with the following properties:\\ 
163~kpc separation/1.3~mag difference; 155~kpc separation/4.4~mag difference.\\ 

\noindent
The first of these is a very bright satellite, the other much fainter
but just an SMC-analogue in mag difference; both are very distant from
the central galaxy, but do lie close to one another, 23~kpc in
projected separation.  Figure~\ref{fig:n4675} shows just the two
satellites, not NGC~4675; the brighter of the two appears to be a
disturbed barred spiral galaxy, consistent with the classification of
it as being too bright to be comparable with the MC.  The fainter
satellite shows very intense, centrally concentrated H$\alpha$
emission.\\

\noindent
{\bf NGC~4708}\\ 
2 satellites found, with the following properties:\\ 
101~kpc separation/7.2~mag difference; 191~kpc separation/5.4~mag difference.\\ 

\noindent
Both are very faint and distant satellites, which are also well
separated from one another (152~kpc).\\

\noindent
{\bf NGC~5303}\\ 
2 satellites found, with the following properties:\\ 
20~kpc separation/1.87~mag difference; 33~kpc separation/3.78~mag difference.\\

\noindent
Individually these are good LMC and SMC analogues in mag difference
and proximity to their host.  However, they do not appear to be a
satellite pair, with a larger satellite-satellite separation (39~kpc)
than the host-satellite separations. Morphologically, the brighter
satellite appears to be an edge-on disk, possibly with a bulge, while
the potentially `SMC-like' satellite has a cometary appearance in red
continuum and a single compact region of H$\alpha$ emission (see
Fig.~\ref{fig:n5303}).

\vspace{0.3 cm} Overall we find 5 cases of pairs of satellites that
are significantly closer to one another than they are to their central
galaxy.  Three of these pairs have projected separations of 41 - 42~kpc,
substantially greater than the current LMC - SMC distance.  The
closest pair, at 13~kpc projected separation, are the two innermost
satellites of NGC~3074.  Probably the best candidates for a bound pair
are the two objects associated with NGC~4675 (shown in
Fig.~\ref{fig:n4675}), with a projected separation of only 23~kpc,
while the mean distance from the pair to the central galaxy is almost
7 times as large.  One question regarding this putative pair is
whether they might be regarded as a central-satellite system in their
own right, given that the brighter galaxy of the pair is one of the 3
`very bright' satellites, substantially more luminous than the LMC.

\section{Discussion}

To the best of our knowledge, the only previous search specifically
for star-forming satellites that has been reported in the literature
is our own study using the H$\alpha$GS data, \citet{jame08c}.  Given
that identical methods were used, it is not surprising that the
present study confirms the main result of \citet{jame08c}; the large
majority of central galaxies do not host star-forming satellites.
This conclusion is unlikely to be changed by the levels of
incompleteness identified in section~\ref{ssec:datred} above. The
present study does, however, find larger numbers of satellites per
central galaxy searched than the ratio of 1 satellite per 13 central
galaxies found by \citet{jame08c}; this may be largely to do with the
higher typical luminosities of the central galaxies surveyed in the
present study. To summarise our numbers for comparison with other
studies, Fig.~\ref{fig:sepmagdiff} shows 1 bright, 6 LMC-like, 15
SMC-like and 15 faint satellites within the `Magellanic radius', from
a total sample of 143 central galaxies searched.  To correct for
incompleteness in the searched volume, these observed numbers should
be multiplied by 1.35.  Extending this to all galactocentric radii,
Fig.~\ref{fig:sepmagdiff} shows 3 bright, 9 LMC-like, 22 SMC-like and
28 faint satellites, which should be multiplied by 1.96 to correct for
volume incompleteness.

\citet{holm69} searched for satellites around 174 galaxies out to a
galactocentric radius of $\sim$57~kpc (after correction to H$_0$=70).
He found 274 `physical' companions (i.e. after a statistical
background subtraction).  For a subset of 53 central galaxies at
Galactic latitude $>$30$^{\circ}$, he found 82 such companions with a
visual absolute magnitude estimated to be brighter than --15.3.  This
corresponds approximately to our SMC-like limit, and within this
radius we find only 11 star-forming satellites, at least an order of
magnitude fewer than the total satellite numbers inferred by
\citet{holm69}.  Similarly, \citet{lorr94} find 1 satellite brighter
than M$_B$=--16.5 (corresponding to a luminosity of
6.2$\times$10$^8$~L$_{\odot}$, somewhat brighter than the SMC) per
central galaxy.  We find only 15 satellites this bright in our sample, 
or 29 after correction for volume incompleteness, still only
1 star-forming satellite per 5 central galaxies searched. Our low
numbers compared with both \citet{holm69} and \citet{lorr94}
may imply efficient truncation of SF in satellites, such that
the majority of satellites become red, quiescent systems well
before they merge with their central galaxies, or there may be 
some remaining inclusion of line-of-sight projected companions 
in the earlier studies.

It is also of interest to compare with \citet{zari93}, who searched 45
central galaxies of types Sb - Sc, with absolute magnitudes M$_B$
--19.5 to --20.5.  They found 69 satellites, with a distribution of
relative luminosities that is generally consistent with that of the
present sample (see Section~\ref{sec:res}), and concluded that
close-in satellites have marginally smaller sizes than those at large
galactocentric distance. There is some evidence for a similar effect
in Fig.~\ref{fig:sepmagdiff}, in the sense that the faintest
satellites seem to be preferentially close-in, with the upper
right-hand corner of the plot being underpopulated.

Many satellite studies have made use of the SDSS, exploiting the
huge numbers of galaxies and the immense statistical weight this
provides.  However, it should be noted that the SDSS studies explore
an almost completely distinct part of parameter space from the present
work, in particular with regard to the luminosity of satellites found.
For example, \citet{ann08} studied a sample of 2254 central galaxies
from SDSS Data Release 5, with mean luminosities close to $L_{\star}$
and thus very comparable to the present sample. They found 4986
companions associated with these galaxies, and studied the effects of
central galaxy type and proximity on the properties (early-
vs. late-type) of the companions.  However, it is important to note
that these companions have a median luminosity difference of only 1.8
mag relative to their central galaxies. Our study finds much fainter
satellite galaxies, with almost half being less than 1\% of the
luminosity of their central galaxy.  

We now turn to a comparison with theoretical studies, and in
particular with the recent predictions from $\Lambda$CDM models.  The
main point to make here is that the MW appears unusual in the
possession of bright, star-forming satellites, and thus the prediction
of such satellites \citep{libe07} should not be taken as validation of
models.  Thus the modifications to models suggested by \citet{simo07},
\citet{kopo09} and \citet{krav10} to produce the full range of MW
satellites including the MC are not required in order to explain the
satellite properties of the ensemble of bright disk galaxies studied
here.  More positively, the finding of \citet{bens02} (reproduced in
the `baseline' versions of many subsequent simulations), that fewer than
5\% of disk galaxies have satellites resembling the LMC, is consistent
with our results.

This scarcity of satellites found here argues against such systems as
a dominant source of gas supply at the current epoch.  Most central
galaxies have no gas-rich satellites within the volumes surveyed
here. If such satellites are present but typically outside the
surveyed volume, this would put them at galactocentric distances
greater than $\sim$200~kpc. The typical dynamical friction timescales
would then be at least several Gyr \citep{boyl08,taff03}, clearly too
long for substantial gas supply for a large spiral, which requires the
gas content of a Magellanic-type satellite every Gyr or so. Fainter
satellites have even longer dynamical timescales and so seem unlikely
to provide a significant contribution to the gas supply, even if there
is a strong upturn in numbers faintward of our detection limits.

We repeat the major caveat on the present work, which is that this
technique is completely insensitive to gas-poor red-sequence
satellites.  A survey of satellite galaxies of all types (M. Prescott
et al. in preparation) is currently being prepared from the Galaxy And
Mass Assembly (GAMA) project \citep{driv09,bald10}, using the AAOmega
spectrometer on the Anglo-Australian Telescope, and will provide a
useful complementary view of satellite populations.

This work has provided a number of satellites which can be used to
compare the detailed properties of the LMC and SMC, with several
examples of analogues for each.  The one aspect of the Clouds for
which there is no clear analogue is their binary nature, with all of
the possible satellite pairs being either significantly more widely
separated than the MC, or very close to the central galaxy.  This is
an interesting discovery in itself, but is also somewhat
disappointing, as there are important questions on the role of
binarity in forming the Magellanic Stream, which could have been
clarified by observing more such examples.

\section{Conclusions}

Our main conclusions are as follows.  Magellanic-type satellite
galaxies are rare; for the majority (approximately two-thirds) of the
central galaxies searched, no star-forming satellites at all have been
found. Among those SF satellites that are found, the MC are
significantly brighter than the average, closer to their central
galaxy than most, and are also unusual in that they form a close pair.
We caution against using the MW and its immediate environment as a
model for simulations of galaxy formation, particularly as regards
bright satellites.  However, some recent models that are not tuned to
produce MC-like satellites do accurately predict the bright satellite
fractions found in the present study.  Our results also highlight
substantial problems with gas-rich satellites as the major source of
gas supply to maintain SF in disk galaxies.  The major caveat on this
work is that it applies to star-forming satellites only, so studies of
the red satellite fraction, e.g. from the GAMA survey, are essential
to provide a full picture of the satellite population.

The next paper in this series will provide a full catalogue of satellite
properties, including the 62 discussed here, resulting from a larger and 
more eclectic sample of H$\alpha$ imaging, including $R$-band and H$\alpha$
luminosity functions, stellar masses, SF rates,
and the effects of central galaxy proximity on SF properties.

\section*{Acknowledgments}

This research has made use of the NASA/IPAC Extragalactic Database
(NED) which is operated by the Jet Propulsion Laboratory, California
Institute of Technology under contract with the National Aeronautics
and Space Administration.  The Isaac Newton Telescope is operated on
the island of La Palma by the Isaac Newton Group in the Spanish
Observatorio del Roque de los Muchachos of the Instituto de
Astrof\'isica de Canarias.  PAJ and CFI are happy to acknowledge the
UK Science and Technology Facilities Council for research grant and
research studentship support, respectively. We thank Sue Percival for
useful comments on a draft of this paper, and the referee, Hong Bae
Ann, for a positive and helpful report.



\label{lastpage}

\end{document}